\documentclass[preprint]{aastex63}

\usepackage{lineno}

\newcommand{\N}[1]{\ensuremath{{{N}[{#1}]}}\xspace}

\newcommand{\kms}{\ensuremath{\mathrm{km\,s^{-1}}}}

\newcommand{\A}[1]{\ensuremath{A_{\mathrm{#1}}}\xspace}

\newcommand{\nhtwo}{\ensuremath{n_{{\rm H}_2}}\ }

\shorttitle{HCN in the Perseus Cloud}
\shortauthors{Dame \& Lada}
\begin{document}

\title{A Complete HCN Survey of the Perseus Molecular Cloud}

\correspondingauthor{Tom Dame}
\email{tdame@cfa.harvard.edu}

\author[0000-0003-0109-2392]{T. M. Dame}
\affiliation{Center for Astrophysics $|$ Harvard \& Smithsonian \\
60 Garden Street \\
Cambridge, MA 02138, USA}

\author[0000-0002-4658-7017]{Charles J. Lada}
\affiliation{Center for Astrophysics $|$ Harvard \& Smithsonian \\
60 Garden Street \\
Cambridge, MA 02138, USA}

\begin{abstract}

We present a survey of the Perseus molecular cloud in the J $=$ 1$\rightarrow$0 transition of HCN, a widely used tracer of dense molecular gas. The survey was conducted with the CfA 1.2 m telescope, which at 89 GHz has a beam width of 11' and a spectral resolution of 0.85 km s$^{-1}$. A total of 8.1 deg$^2$ was surveyed on a uniform 10' grid to a sensitivity of 14 mK per channel. The survey was compared with similar surveys of CO and dust in order to study and calibrate the HCN line as a dense gas tracer. We find the HCN emission to extend over a considerable fraction of the cloud. We show that the HCN intensity remains linear with H$_2$ column density well into the regime where the CO line saturates. We use radiative transfer modeling to show that this likely results from subthermal excitation of HCN in a cloud where the column and volume densities of H$_2$ are positively correlated. To match our HCN observations the model requires an exponential decrease in HCN abundance with increasing extinction, consistent with HCN depletion onto grains. The modeling also reveals that the mean volume density of H$_2$ in the HCN emitting regions is $\sim$ 10$^4$ cm$^{-3}$, well below the HCN critical density. For the first time, we obtain a direct measurement of the ratio of dense gas mass to HCN luminosity for an entire nearby molecular cloud: $\alpha$(HCN) $=$ 92 M$_\odot$/(K km s$^{-1}$ pc$^2$).   

\end{abstract}


\section{Introduction} \label{sec:Introduction}

The lowest rotational transition of CO at 115 GHz is widely acknowledged to be the best general-purpose tracer of the total mass, distribution, and kinematics of the largely unseen H$_2$ gas in molecular clouds and it is for this reason that the CfA 1.2 meter telescope has mainly remained tuned to CO for over four decades. Surveys by this and other telescopes have taught us that star formation takes place almost exclusively within molecular clouds, but surprisingly the rate of star formation in a molecular cloud seems to depend much less on its total mass than on the mass of gas it contains in the form of dense cores and filaments (Lada et al. 2010). It has long been appreciated that star formation occurs within such dense regions (e.g., Lada 1992), but their high opacity in CO and limited spatial extents make them inconspicuous in CO surveys. 
Instead, in Galactic clouds, rarer molecular species, such as CS, HCN, NH$_3$, and N$_2$H$^+$, with higher dipole moments than CO, have traditionally been used to trace the densest gas in the  clouds (e.g., Martin \& Barrett 1978, Benson \& Myers 1989, Jackson et al. 2013, Pety et al. 2017 and many others) since such molecules are not as easily collisionally excited to a state of local thermodynamic equilibrium as CO. However, complexities of molecular excitation, radiative transfer, and chemistry have long rendered direct, quantitative interpretation of such subthermal molecular line emission problematic.  

As a result, measurements of infrared dust extinction (e.g., Lada et al. 2010, 2012) and dust continuum emission (e.g., Lombardi et al. 2014, Zari et al. 2016) have been used to more effectively trace the dense component of molecular clouds.   After H$_2$, dust is the second most abundant constituent of molecular clouds and is well mixed with the H$_2$ gas. Because measurements of the dust are not hampered by the factors (e.g., excitation, opacity, and chemistry)  that complicate the interpretation of molecular line observations, dust provides more direct and reliable measurements of the total column densities and masses of molecular clouds. Unfortunately, current observational capabilities largely limit the use of dust as a dense gas tracer in distant clouds and external galaxies and observations of molecular lines still remain the best potential tracers of dense molecular material outside the Galaxy. To increase the effectiveness of these molecular-line tracers it is essential to calibrate them with dust measurements wherever possible. 

After CO and its isotopologues, HCN is among the most readily detected molecular species at millimeter wavelengths with the potential to trace dense gas in molecular clouds. Knowledge of the dense gas contents of molecular clouds is crucial for understanding star formation and galaxy evolution. This is because local GMCs in the Milky Way are described by a star formation law in which their star formation rates (SFRs) are both tightly and linearly correlated with the masses of high column density  gas in the clouds (Lada et al. 2010, 2012).  A similar linear star formation law also appears to characterize entire galaxies where their total infrared luminosities (a presumed proxy for SFRs)  are tightly correlated with their HCN luminosities (Gao \& Solomon 2004, Shimajiri et al. 2017, Jim{\'e}nez-Donaire et al. 2019). How are these two star formation laws physically connected? Clearly, calibrated HCN observations are of high astrophysical significance for ultimately understanding the nature of this connection. 

In their widely-cited paper, Gao \& Solomon (2004) emphasized the need for ``systematic and unbiased large-scale observations'' of HCN in Galactic molecular clouds in order to calibrate HCN as a dense gas tracer and to better understand the physical nature of the tight SFR-dense gas correlation. Yet to date, owing primarily to the weakness of HCN emission relative to that of CO, no more than one square degree of sky has been systematically surveyed in HCN emission in any nearby cloud (e.g. Pety et al. 2017, Shimajiri et al. 2017). Recently, Tafalla et al. (2021) mapped HCN (and other molecular species) in Perseus using a stratified random sampling technique in which ten points were randomly sampled at each of ten logarithmically-spaced levels of column density. As we discuss below, this approach can be very effective at characterizing the radiative transfer of molecular lines, but such a severely under sampled study cannot be used to measure the total HCN luminosity of Perseus nor its dense gas conversion factor (hereafter $\alpha({\rm HCN})$). For that, a well-sampled survey is required. 

Here we present the first complete HCN survey of a nearby molecular cloud, carried out with the CfA 1.2 meter telescope in the 1$\rightarrow$0 transition. The primary goals of this survey were to: 1) accurately measure the total HCN luminosity from an entire, spatially resolved, GMC in order to make a  robust determination of $\alpha$(HCN) for molecular clouds, and 2) to determine the range of molecular hydrogen column and volume densities which produce detectable HCN emission. We chose the Perseus cloud because of both its typical level of star formation activity---not as flamboyant as the Orion and Ophiuchus clouds nor as quiescent as the Pipe and California clouds---and its ideal declination for mapping from the latitude of Cambridge. In addition, the Perseus cloud has been fully mapped in both dust emission by Herschel (Andre et al. 2010) and Planck (Planck Collaboration et al. 2014) and in near infrared extinction (Lombardi et al. 2010), providing a well-calibrated, high-resolution inventory of its dust content (Zari et al. 2016).

\begin{figure}
\epsscale{0.6}
\plotone{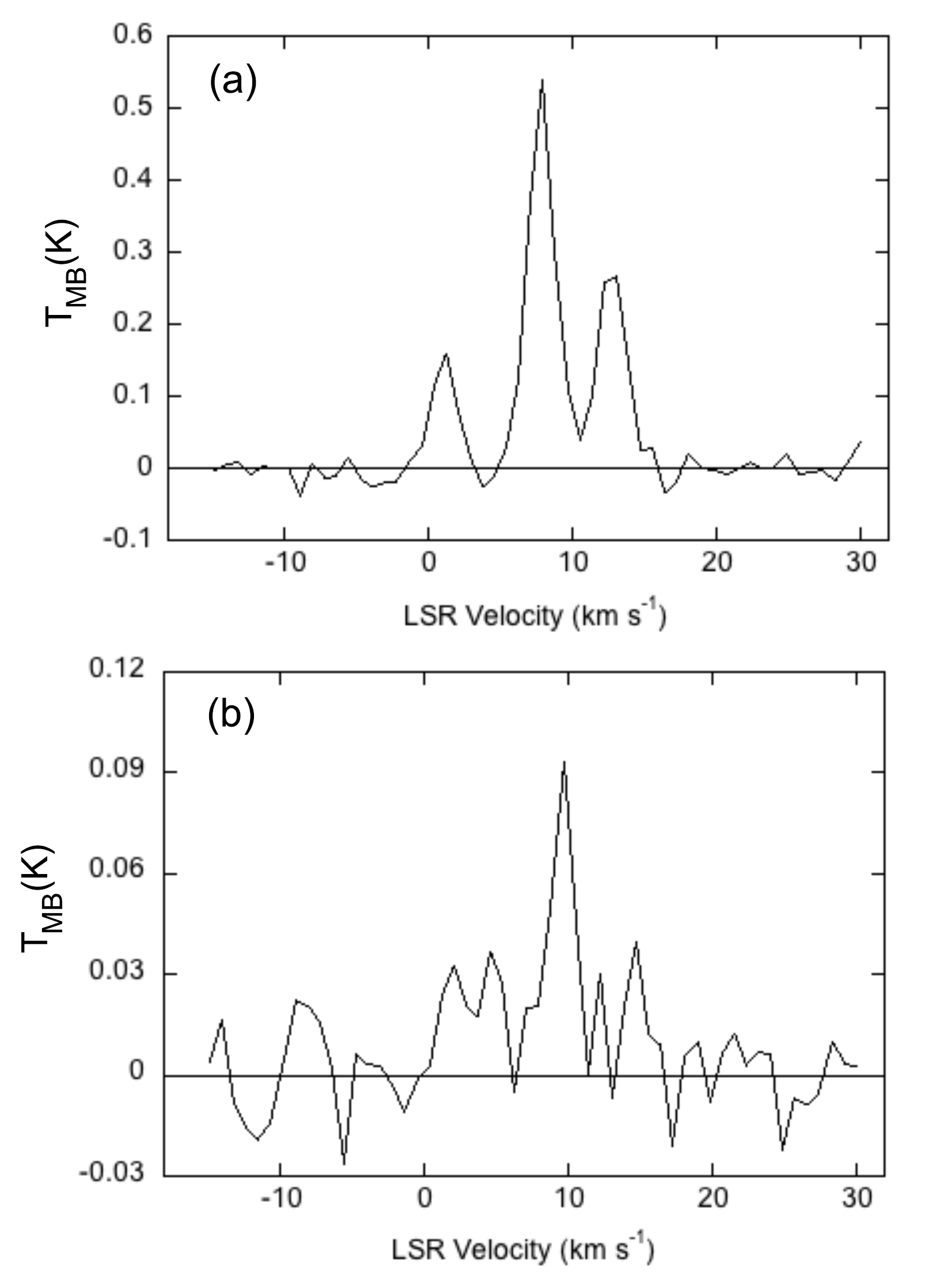}
\caption{(a) HCN J=1-0 spectrum at the strongest point in the survey, $l$ = 158.333, $b$ = -20.5. (b) HCN spectrum at a typical position, $l$ = 160.0, $b$ = -19.167. }
\end{figure}
  
In the next section we describe our HCN survey in detail and show sample spectra. In Section 3 we compare our HCN map to a previous CO map of Perseus obtained with the same telescope and to the high resolution dust column density map of Zari et al. (2016). The latter comparison is used to provide an empirical conversion factor between HCN luminosity and total dense gas mass in the Perseus cloud. In Section 4 we use our HCN survey, the Zari et al. dust map, and radiative transfer modeling to investigate both the relation between HCN emission and H$_2$ column density and the range of gas volume densities traced by HCN. In Section 5 we summarize our results.

\section{Observations}

The HCN ($J = 1\rightarrow0$) survey was conducted between February and April 2019 with the CfA 1.2 meter telescope, which is fully described in Dame et al. (2001). Observations in the LSR frame were centered at a frequency of 88631.847 MHz, corresponding to the center line in a group of 3 HCN hyperfine lines ($F = 1-0, 1-0, 2-1$); the other lines are offset by $+$4.82 \kms\  and -7.07 \kms\  in radial velocity from the central component. A 256 channel filter bank spectrometer with 0.25 MHz channels provided a velocity resolution of 0.85 \kms\  and coverage over 216 \kms. We removed instrumental baseline structure by frequency switching every 1 s by 10 MHz (34 \kms) and subtracting the data from the two phases. We subsequently removed a 5th order polynomial and folded the spectra by averaging each spectral channel with its corresponding reference-phase mate 10 MHz higher in frequency. Figure 1 shows two of the final spectra from the survey, one at the peak position and one at a typical position in the cloud.

Although the intensity of the HCN lines are typically 50-100 times weaker than those of CO, the sky noise at 89 GHz is roughly 3 times lower than at 115 GHz, owing mainly to a larger offset from a strong O$_2$ telluric  absorption line at 118 GHz. At a typical observing elevation of 40$^{\circ}$, the total system noise (sky plus receiver plus side lobe leakage) was $\sim$240 K at the HCN (J $=$ 1$\rightarrow$0) frequency, compared with $\sim$550 K at CO (J $=$ 1$\rightarrow$0). The total integration per sky position was $\sim$40 minutes, which yielded a uniform rms noise of 14 mK per channel. We note that the integration times were not fixed but rather were adjusted for each observation based on the instantaneous system noise to achieve that target rms. For various practical reasons including calibration stability, we actually observed the entire map 5 times over with an average integration time of $\sim$8 minutes per point in each map. 

Based on a measured beam width of 8.4' at 115.3 GHz (see Appendix A in Dame et al. 2001), the beam at 89 GHz is estimated to be 10.9' (HPBW). Our observations were guided by an existing 1.2 m CO survey of the Perseus cloud from Dame et al. (2001), spatially smoothed to an angular resolution of 10.9'. As shown in Figure 2b, we searched for HCN at all positions where the integrated integrated intensity of the CO line, W(CO), was in excess of 13 K km s$^{-1}$, corresponding to an expected W(HCN)\footnote{Throughout this paper W(HCN) corresponds to the intensity integrated over the three main hyperfine components. The expected W(HCN) assumes W(CO)/W(HCN) = 71, a value we derived by comparing CO (Wilson et al. 2005) and HCN (Pety et al. 2017) in Orion B. The results presented below confirm that this ratio is approximately correct.} in excess of 3 times the target instrumental noise. A total of 292 positions were observed on a 10' grid in Galactic longitude and latitude, covering a total area of 8.1 deg$^2$.

\vskip 0.25in

\section{Results and Discussion}

\subsection{Comparison of HCN, CO and Dust Extinction}

A map of velocity-integrated HCN intensity derived from our new survey is shown in Figure 2a, above a comparable CO map from the CfA survey in Figure 2b and a map of K band dust extinction from Zari et al. in Figure 2c. The latter two maps have been spatially smoothed to the intrinsic 11' resolution of the HCN map. The HCN surveyed area is outlined on all three maps. It is clear from Figure 2b that our survey area covers most (78\%) of the cloud's CO luminosity, and within that area we detect HCN at the 3-$\sigma$ level in 60\% of the positions observed. Using the dust map in Figure 2c to infer gas mass, we find that the HCN detected positions contain 1.2 $\times$ 10$^4$ M$_\odot$ of gas, corresponding to 75\% of the total gas mass in the survey area and 55\% of all the CO emitting gas in the Perseus cloud. 
 
\begin{figure}
\plotone{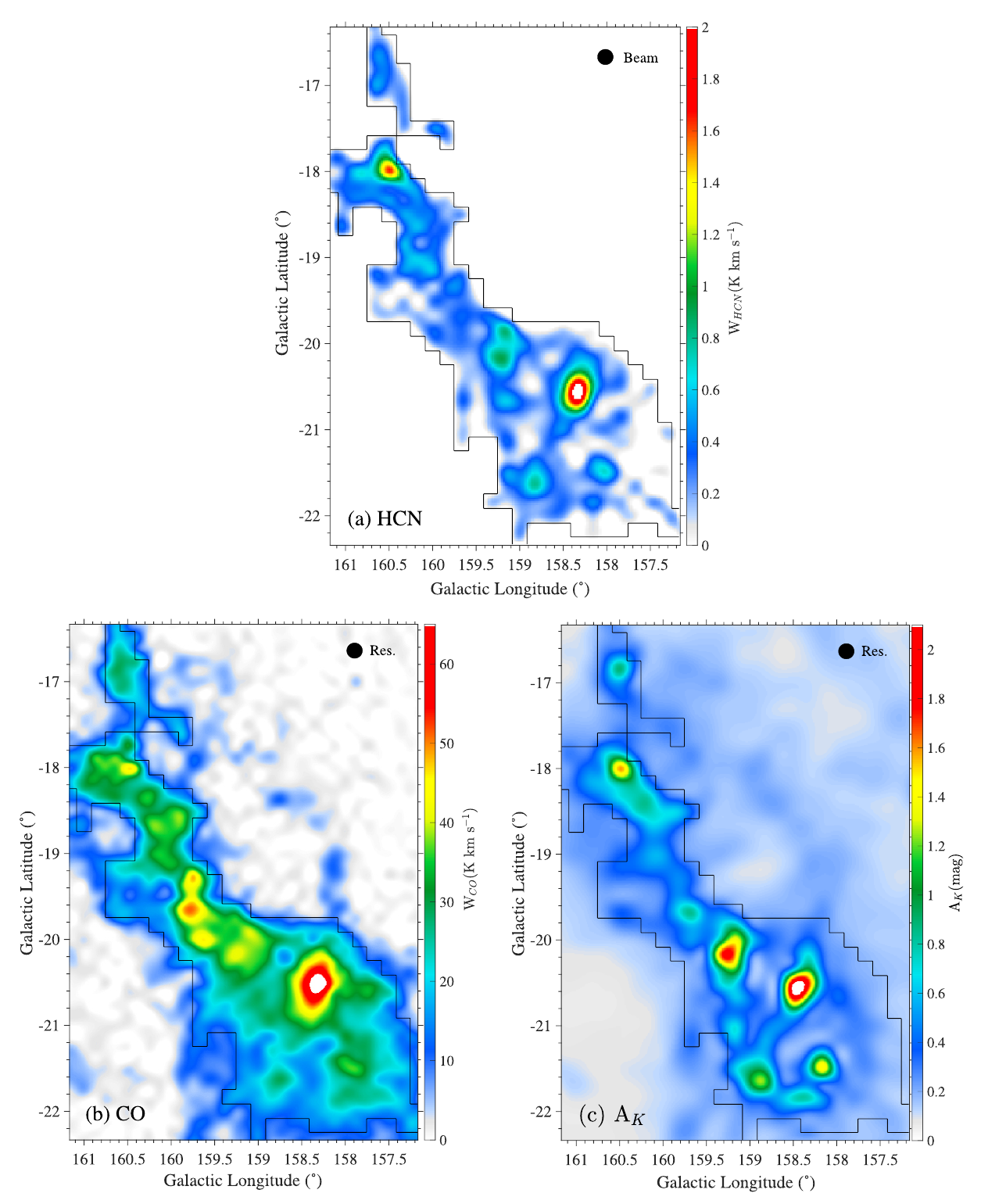}
\caption{(a) HCN emission integrated over the full velocity range in which CO and HCN are detected, -5 to 15 \kms. (b) CO emission integrated over the same velocity range and smoothed to an angular resolution of 11’ to match that of the HCN survey. (c) The A$_K$ map of Zari et al. (2016) likewise smoothed to an angular resolution of 11’. All three maps have the HCN sampling boundary overlaid.  }
\end{figure}

\begin{figure}
\epsscale{0.7}
\plotone{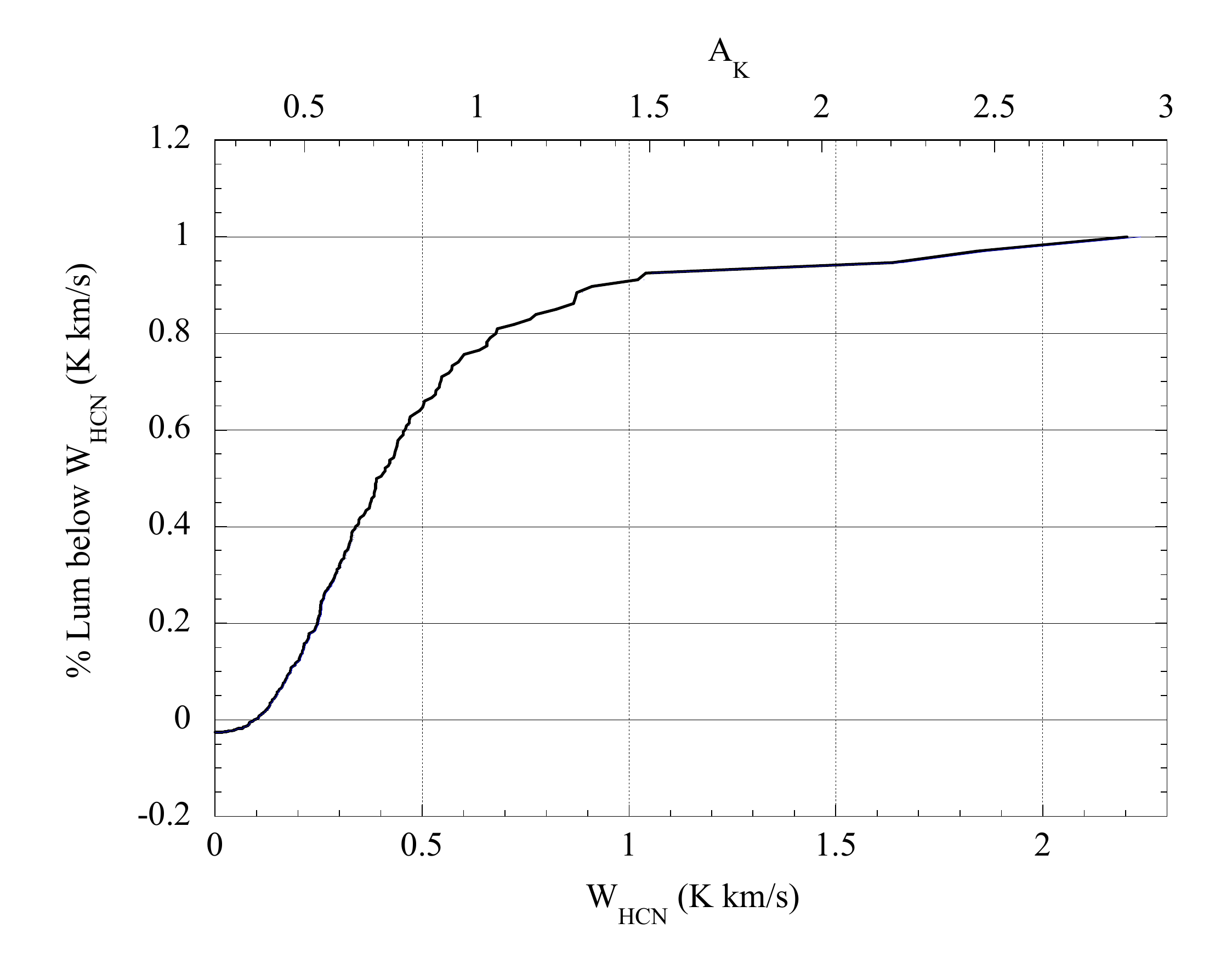}
\caption{\label{fig:fWvsW} The percentage of total cloud luminosity contributed by positions below W(HCN). For example, $\sim$50\% of the total luminosity is contributed by positions with W(HCN) $<$ 0.4 K km s$^{-1}$. Along the top axis we used the linear fit in Figure 4d to convert W(HCN) to an approximate value of A$_K$.   }
\end{figure}

In Figure \ref{fig:fWvsW} we show the cumulative fraction of total HCN luminosity as a function of W(HCN) and extinction. Note that 60\% of the total HCN luminosity is emitted in regions of the cloud \emph{below} the standard dense-gas extinction threshold of A$_K$ $=$ 0.8 mag. This result is both consistent with and reinforces earlier suggestions of very extended HCN emission that were based on more spatially limited HCN surveys within a few nearby clouds (Pety et al. 2017, Shimajiri et al. 2017, Kauffmann et al. 2017, Tafalla et al 2021). It is becoming increasingly clear that comparable wide-field surveys of entire molecular clouds will reveal similarly extensive HCN emission as we find in Perseus. 

\begin{figure}
\plotone{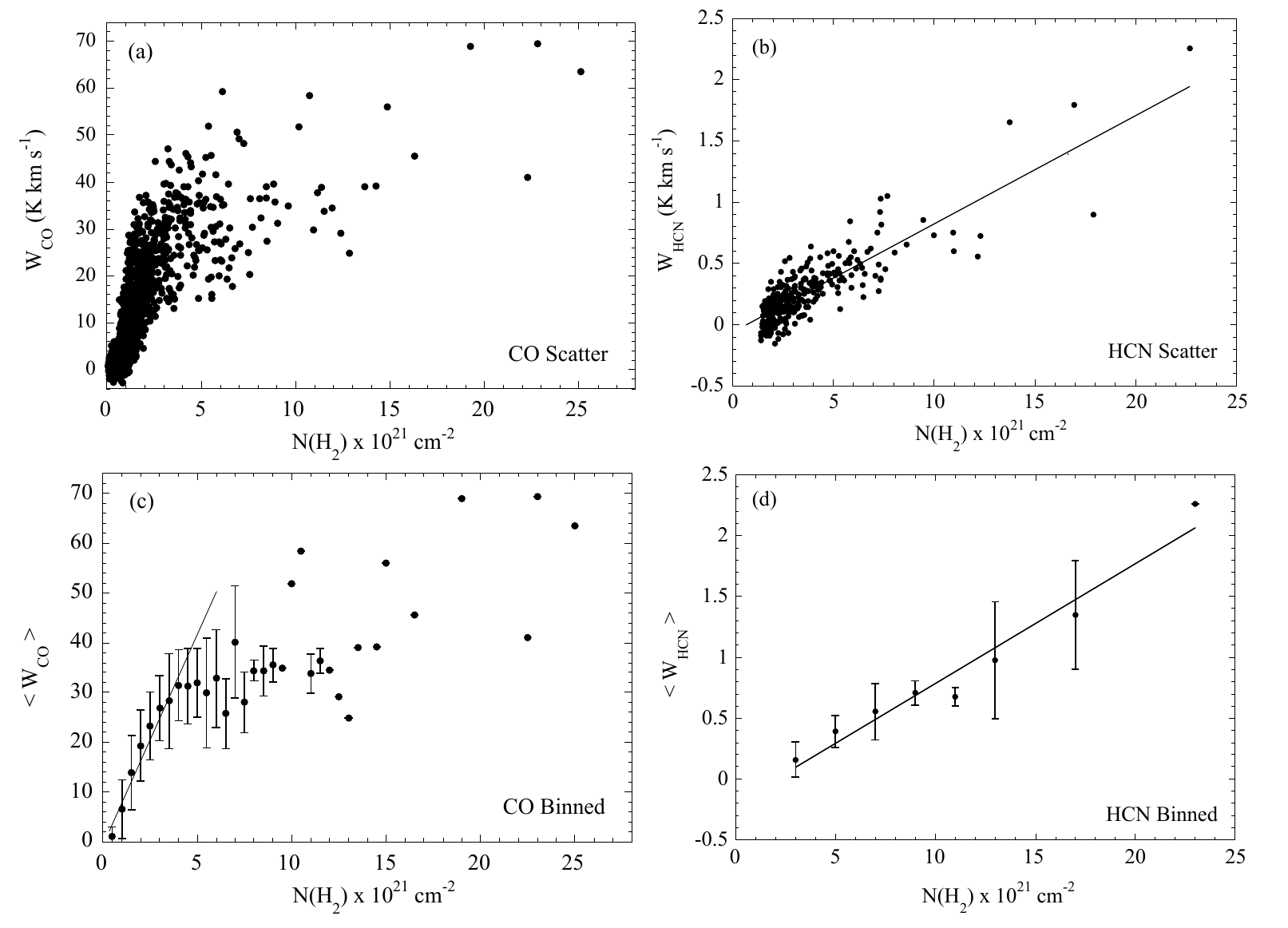}
\caption{Scatter plots derived from the maps in Figure 2, with A$_K$ converted to H$_2$ column density using the gas-to-dust ratio in Zari et al. (2016). (a) Scatter plot derived from the CO and dust maps. (b) Scatter plot derived from the HCN and dust maps. (c) The points from (a) averaged in bins 0.5 x 10$^{21}$ cm$^{-2}$ wide in N(H$_2$). The line was fitted to points at N(H$_2$) $\leq$ 4 x 10$^{21}$ cm$^{-2}$. (d) The points from (b) averaged in bins 3.5 x 10$^{21}$ cm$^{-2}$ wide in N(H$_2$). The line was fitted to all points and implies X$_{HCN}$ $=$ 102 x 10$^{20}$ cm$^{-2}$ K$^{-1}$ km$^{-1}$ s. In (c) and (d), the error bars indicate dispersion about the mean in the bins; no dispersion is indicated for bins with only one data point. For a standard gas-to-dust ratio,  1 A$_V$ $\approx$ 1 x 10$^{21}$ cm$^{-2}$, so the numbers on the x axes approximately represent A$_V$.}
\end{figure}

As noted above, the dust extinction map in Figure 2c can be linearly scaled to a map of N(H$_2$) by adopting a standard gas-to-dust ratio (see Zari et al. for details), and we can use that map to test the optical depth properties of the HCN and CO lines in Perseus. Scatter plots of HCN and CO emission as functions of N(H$_2$) are shown in the top two panels of Figure 4, while the corresponding lower panels average the emissions in bins of N(H$_2$). At N(H$_2$) $<$ 4 x 10$^{21}$ cm$^{-2}$ the CO line does a good job of linearly tracing the molecular gas with a corresponding X factor of 1.2 x 10$^{20}$ cm$^{-1}$ K$^{-1}$ km$^{-1}$ s, while at higher column densities we see the well-known saturation of the CO line. In contrast, the HCN plots in Figure 4 show no signs of saturation, the velocity integrated HCN intensity remaining linear with N(H$_2$) and yielding a corresponding  HCN-to-H$_2$ conversion factor, X$_{HCN}$ of 102 x 10$^{20}$ cm$^{-2}$ K$^{-1}$ km$^{-1}$ s, very close to the value derived by Tafalla et al. (2021). Thus, even though its high dipole moment suggests that, relative to CO, HCN should be a useful tracer of gas at high \emph{volume} density, it is clearly also a good tracer of gas \emph{column} density over the range in N(H$_2$) of 0.2--2.3 x 10$^{22}$ cm$^{-2}$ (i.e., A$_v$ $\approx$ 2-20 mag), at least in Perseus.

\subsection{Calibrating HCN luminosity as a Tracer of Dense Gas Mass}

Although the X-factor is useful for calculating molecular hydrogen column densities from HCN integrated intensities in resolved nearby clouds, beam dilution limits its use in external galaxies (and very distant clouds) where the molecular gas is unresolved. Because for unresolved sources the line flux, or equivalently line luminosity, is conserved with varying beam size, it is more convenient to use a calibrated mass-luminosity relation to directly extract dense gas masses. 

Following Gao \& Solomon (2004), we define the HCN mass-luminosity relation as:
\begin{equation}
\alpha({\rm HCN}) = M_{dg}/L({\rm HCN}) 
\end{equation}
\noindent
where $M_{dg}$ is the mass (in units of M$_\odot$) of the molecular gas at high volume density and $L$(HCN) is the total HCN(1-0) luminosity (in units of K km s$^{-1}$ pc$^2$) of the cloud.  

Knowledge of the value of $\alpha$(HCN) has been sparse, and until recently there were no directly calibrated measurements. Gao \& Solomon (2004) were the first to estimate a value for $\alpha$(HCN). From theoretical considerations they indirectly determined a value of 10 by assuming virial equilibrium, a uniform density of \nhtwo = 3 $\times 10^4$ cm$^{-3}$, and a line brightness temperature of 35 K for the HCN emitting gas. Among the first attempts to empirically determine $\alpha$(HCN) in the Milky Way was that of Wu et al. (2010), who obtained HCN measurements toward a sample of massive, dense cores. Similar to Gao \& Solomon, they used the virial theorem to estimate masses of the HCN emitting gas and derived a value of $\alpha$(HCN) = 20 for the dense cores. The availability of extensive and complete surveys of dust extinction and emission in GMCs (e.g., Lombardi et al. 2006, 2010, 2011, 2014) along with limited HCN mapping (e.g., Pety et al. 2017; Shimajiri et al. 2017, Kauffmann et al 2017) have recently enabled more direct measurements of $\alpha$(HCN) by using high extinction as a proxy for high volume density. Using {\sl Herschel} dust maps, Shimajiri et al. (2017) derived values for $\alpha$(HCN) in the range  35--454 in ten heavily extincted (A$_V$ $\gtrsim$ 8) portions of three nearby GMCs. Collectively, however, these regions covered a total of only 0.65 deg$^2$, a small fraction the total cloud extents ($\sim$ 10 deg$^2$). To facilitate a more appropriate comparison to extragalactic observations where global cloud measurements will also include HCN emitted at lower extinctions/densities, Shimajiri et al. (2017) argued that their measurements of $\alpha$(HCN) should be scaled down by a factor of 10, roughly the ratio of gas at A$_V$ $>$ 2 to that at A$_V$ $>$ 8 in typical GMCs. Kauffmann et al.(2017) used infrared extinction maps in a section of the Orion A GMC to measure the dense gas mass (at extinctions A$_V$ $>$ 7) and together with their estimates of L$_{HCN}$ for that region found $\alpha$(HCN) $\leq$ 20 after making approximate adjustments to account for HCN emission in gas with lower extinctions. In the end these studies find results similar to that of Gao \& Solomon but only after uncertain corrections for unobserved HCN gas are included. 

\begin{figure}
\plotone{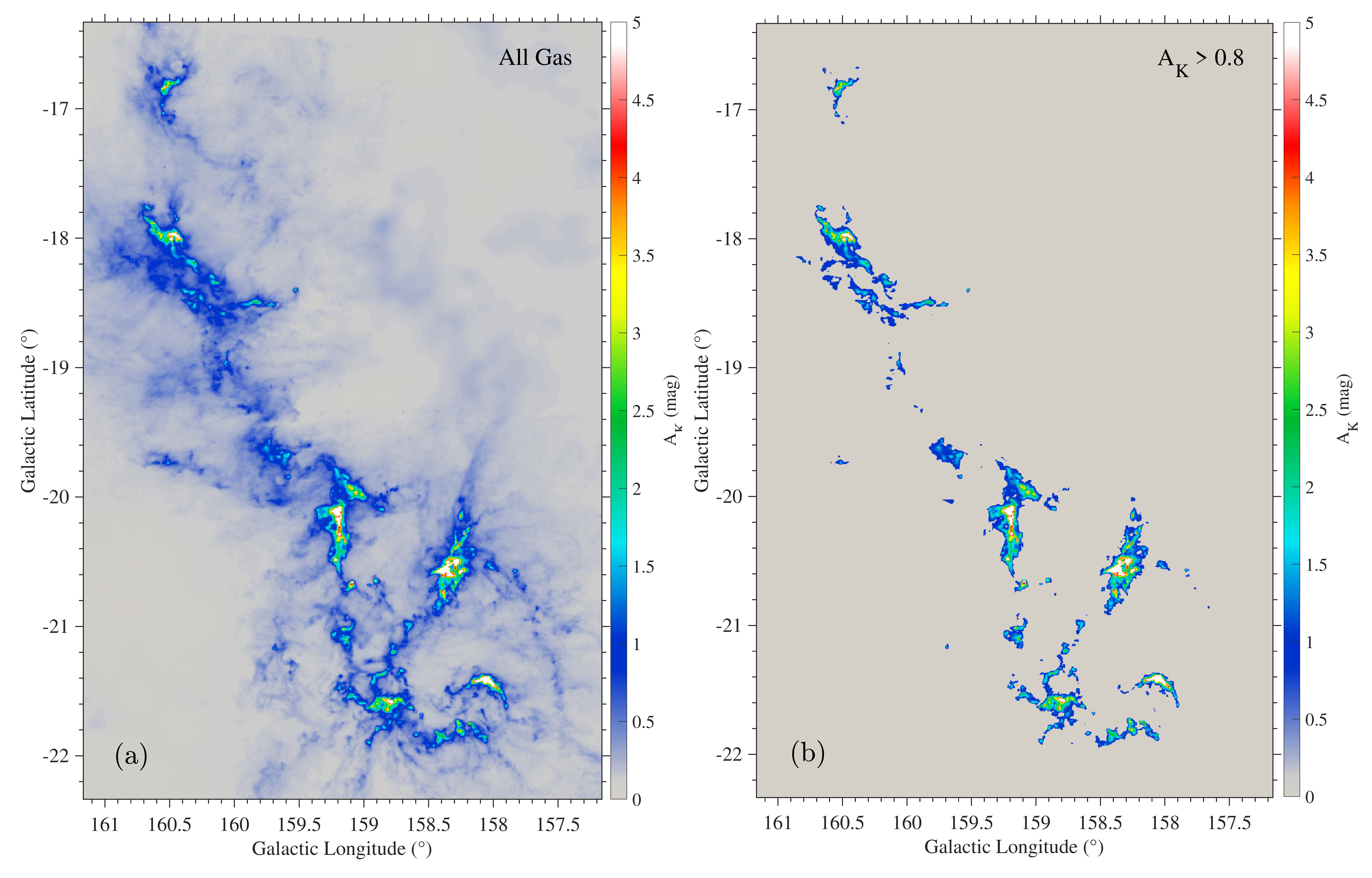}
\caption{ (a) Map of K band extinction from Zari et al. (2016). (b) Same, except only emission at A$_K$ $>$ 0.8 mag is shown. }
\end{figure}

With the data presented here we can, for the first time, directly calibrate $\alpha$(HCN) for an entire nearby molecular cloud. By simply summing over all pixels in our HCN map (Figure 2a) and multiplying by the square of the cloud's distance we obtain L(HCN) = 55.3 K km s$^{-1}$ pc$^2$. The distance to Perseus has recently been measured to high accuracy by Zucker et al. (2019) based on Gaia DR2 data (294 $\pm$ 15 pc). The rms noise in the HCN spectra propagates to an uncertainty on L(HCN) of only 1\%, so a survey with higher angular resolution or sensitivity would scarcely improve on our value. We estimated the amount of HCN luminosity that might lie outside our sampling boundary by scaling the amount of CO emission seen there (Figure 2b) by the mean HCN/CO intensity ratio, which we measured to be $\sim $0.008 near the edges of the cloud. Such weak, unobserved HCN might increase L(HCN) by $\sim $21\%, to 66.9 K km s$^{-1}$ pc$^2$, but this is probably an upper limit, since the HCN/CO ratio likely falls---perhaps even to zero---toward the cloud edges. As an aside, we note that all of the HCN lines that we estimate might lie outside our sampling boundary would not be detectable at our current sensitivity.  

To identify regions of high gas volume density, we adopt a dust column density threshold of A$_K$ $>$ 0.8 mag because, as previously mentioned, the gas mass above that threshold has been shown to be most closely correlated with the star formation rates in local clouds (i.e., Lada et al. 2010; Evans et al. 2014; Heiderman et al. 2010). 
 The premise that our extinction threshold also corresponds to a high volume density threshold follows naturally from the fact that molecular clouds are filamentary in geometry rather than sheet-like, with inwardly increasing column density and volume density gradients. This is perhaps best demonstrated by recent 3D extinction maps of the outer regions of local clouds (Zucker et al. 2021). This idea also derives support from early models of filamentary clouds showing that radial column density gradients likely correspond to radial volume density gradients (e.g. Lada, Alves \& Lada 1999, Bergin et al. 2001). Indeed, Bergin et al. (2001; see their Figure 10) derived a relation between n(H$_2$) and visual extinction in the IC 5146 cloud indicating that n(H$_2$) $>$ 10$^4$ cm$^{-3}$ corresponded to  Av $>$ 7 mag (i.e., A$_K$ $ >$ 0.8 mag)\footnote{We note that A$_V$ is related to A$_K$ through the reddening law: A$_K$ $=$ 0.112 A$_V$ (Rieke \& Lebofsky 1985). } for that cloud.  More recently a number of different turbulent MHD simulations of cloud formation and evolution made over a wide variety of physical scales were all found to show clear correlations between extinction and hydrogen volume density. These derived correlations are quantitatively very similar to each other (Bisbas et al. 2019 and references therein; see their Figure B1) and to the early cloud modeling mentioned above.  These various models and simulations consistently find that A$_K$ $>$ 0.8 mag corresponds to n(H$_2$) $>$ 10$^4$ cm$^{-3}$. 

The dense gas mass of the Perseus cloud can be derived from the map of dust extinction shown in Figure 5. Adopting the gas-to-dust ratio given in Zari et al. and defining dense gas as that characterized by extinctions A$_K$ $>$ 0.8 mag, we derive from Figure 5b a dense gas mass of $M_{dg} = 0.51 \times 10^4$ M$_\odot$ and $\alpha$(HCN) $=$ 92 M$_\odot$ $(\rm { K\ km\ s^{-1} pc^2})^{-1}$. Accounting for HCN outside our sampling boundary might decrease this value to 76 M$_\odot$ $(\rm { K\ km\ s^{-1} pc^2})^{-1}$. In any case, we find a value of $\alpha$(HCN) in Perseus that is substantially higher than the value of 10 estimated by Gao \& Solomon. However, assuming virialized clouds, Gao \& Solomon (2004) derived the value of 10 for the dense gas conversion factor using the equation $\alpha$(HCN) $=$ 2.1 $\times$ n(H$_2$)$^{0.5}$(T$_b$)$^{-1}$ where T$_b$ is the intrinsic brightness temperature of the HCN line. They assumed T$_b$ $=$ 35 K and n(H$_2$) $=$ 3$\times 10^4$ cm$^{-3}$ to get the value of 10. Using values more appropriate for Galactic GMCs (e.g., T$_b$ $\lesssim$ 1 K and n(H$_2$) $\sim$   10$^4$ cm$^{-3}$) would give $\alpha$(HCN) $\gtrsim$ 210 using Gao \& Solomon's formula.

It's worth noting here that another value of $\alpha$ can be calculated directly from the value of X(HCN) derived in the previous section. In fact, for any molecular tracer T, it can be shown that $\alpha$(T)$_X$ $=$ 2.15 $\times 10^{-20}$ X(T). For Perseus this yields $\alpha$(HCN)$_{X}$ $=$ 215. This value is larger than the dense-gas alpha derived above because it applies to \emph{all} of the HCN-luminous gas, not just the dense gas at A$_K$ $ >$ 0.8 mag. Multiplying $\alpha$(HCN)$_{X}$ by L(HCN) yields an HCN-luminous mass of 1.2 $\times10^4$ M$_\odot$, in agreement with our direct measurement of this quantity in Section 3.1. 

\vskip 0.25in

\section{The Column and Volume Densities Traced by HCN} 

The HCN molecule has long been considered a dense gas tracer. Yet, in the Perseus cloud both our observations and those of Tafalla et al. (2021) show that HCN is a good tracer of hydrogen column density as well.  This may seem surprising given that the HCN emission is also characterized by very high optical depths ($\tau$ $\sim$ 2-25) under the conditions typical of this and other clouds.  Tafalla et al. (2021) posited that  the observed increase in W(HCN) with extinction could be a consequence of the subthermal excitation of the lines coupled with density gradients in the cloud.
Consider that the HCN brightness is given by the standard radiative transfer equation, $T_b = (J(T_{ex}) - J(T_{bg})) \times (1-e^{-\tau})$ which for $\tau >> 1$  is dominated by the first term. In this circumstance the observed brightness will depend on this excitation term, which will increase as $T_{ex}$ increases with volume density. If, in addition, the extinction (or column density) is correlated with the volume density, as we argued earlier, the brightness of the lines and W(HCN) will also increase with extinction (Tafalla et al. 2021). 

\begin{figure}
\epsscale{0.8}
\plotone{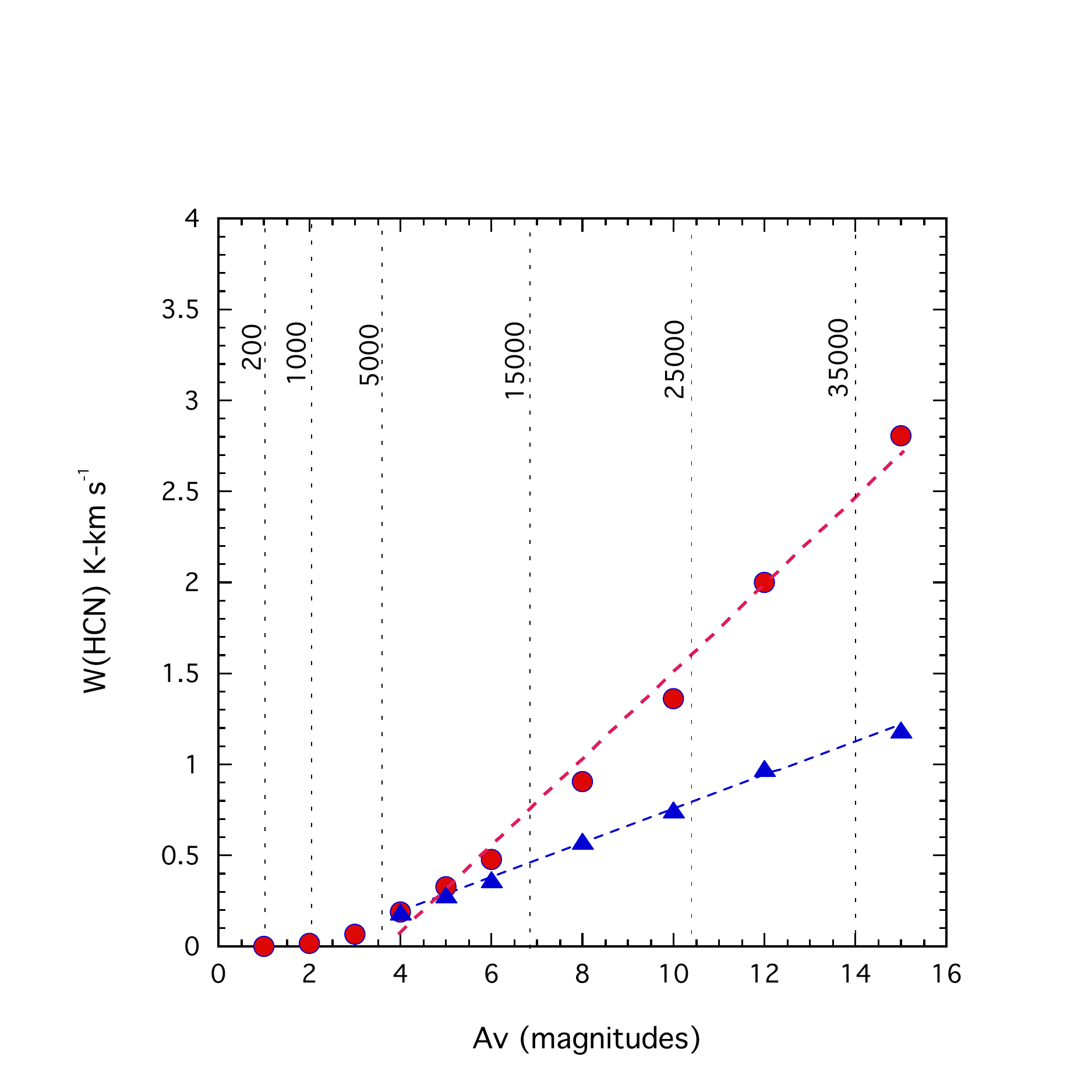}
\caption{\label{fig:WvsAv_model_2km/s} The predicted W(HCN)-A$_V$ relation for a model cloud where column density and volume density are positively correlated. The points are the model predictions for undepleted (circles) and depleted (triangles) gas and the dashed lines are linear fits to the respective points above A$_V$ $=$ 4 mag. Vertical dashed lines correspond to values of n(H$_2$) in units of cm$^{-3}$. These models demonstrate that optically thick HCN emission can effectively trace the total column density in a cloud provided the volume and column density structures of the cloud are correlated.    }  
\end{figure}

To test this idea we used the radiative transfer program RADEX (van der Tak et al. 2007) to predict W(HCN) as a function of A$_V$.  In brief, RADEX can calculate W(HCN) for a uniform medium given N(HCN), n(H$_2$), $T_{\rm K}$, and the observed linewidth $\Delta$v.  We adopted $T_{\rm K}$ $=$ 11 K, based on an ammonia study of 193 dense cores in Perseus (Rosolowsky et al. 2008) and $\Delta$v $=$ 2.3 km s$^{-1}$, based on Gaussian fits to the central hyperfine components in several regions with strong lines in our survey. For a set of $A_V$ values between 0 and 15 mag we calculated: (1) the corresponding N(H$_2$) from the gas-to-dust ratio of Zari et al; (2) the corresponding N(HCN) by assuming an HCN abundance relative to H$_2$ of 3.1 $\times 10^{-9}$ (e.g., Tafalla et al. 2021); and (3) the corresponding n(H$_2$) value from the N(H$_2$)-n(H$_2$) relation of Bisbas et al. (2019).

The predicted values of W(HCN) as a function of A$_V$ are shown by the red circles in Figure \ref{fig:WvsAv_model_2km/s}. Although simplistic, the model provides some interesting insights. First, it predicts that HCN lines only reach our 3$\sigma$ detection limit of 0.17 K km s$^{-1}$ at A$_V$ $\approx$ 3.5 mag. Below that extinction HCN emission is predicted to sharply decrease, with W(HCN) falling to values of only 0.0015 to 0.014  K km s$^{-1}$ between  A$_V$ = 1 and 2 mag, respectively. Such low integrated intensities would present a significant challenge for observations with even the most sensitive receivers, especially since the models do not include the expected UV photodissociaton of HCN near the cloud edges (e.g., van Dishoeck \& Black 1988). Second, the model demonstrates that above A$_V$ $=$ 4 mag, the W(HCN)-A$_V$ relation is indeed linear, the straight line fit to those points in Figure 6 having a correlation coefficient of 0.99. We conclude that our model based on the conjecture of Tafalla et al. can explain the curious fact that HCN emission traces total gas column density surprisingly well despite its high opacity.

We note however that the slope of the red line fit to the model points in Figure 6 corresponds to an HCN-to-N(H$_2$) conversion factor of X$_{HCN}$ $=$ 45 $\times$ 10$^{20}$ cm$^{-2}$ K$^{-1}$ km$^{-1}$ s, about a factor of 2 below our observed value (Figure 4d). Moreover above A$_V$ $\approx$ 5 mag the model systematically over predicts W(HCN) with increasing extinction. Tafalla et al (2021) found a similar result in their modelling of HCN emission in Perseus and attributed it to freeze out or depletion of HCN onto grains. To illustrate the possible effects of depletion on our data we calculated a simple extinction-dependent correction to our model. We assumed no depletion below 4 mag and above 4 mag we required the W(HCN)-A$_V$ relation to remain linear with a slope that matched the observed one as closely as possible and then calculate the corresponding values of N(HCN). 

The depletion-corrected model points and linear fit are shown by the blue triangles and blue dashed line in Figure 6.  As expected, with this adjustment the model predictions much more closely match the observed data, yielding an X(HCN) $=$ 107 $\times 10^{20}$ cm$^{-2}$ K$^{-1}$ km$^{-1}$ s. Moreover, the model shows that N(HCN) is very weakly dependent on A$_V$, with N(HCN) $\sim$ A$_V^{0.14}$ resulting in an exponentially decreasing HCN abundance with cloud depth  (i.e., N(HCN)/N(H$_2$) $\sim$ e$^{-({A_V\over{9.8}})}$)  and supporting the depletion hypothesis. At A$_V$ $=$ 15 mag, N(HCN) is only 20\% higher than it is at 4 mag and is a factor of 3 lower than its undepleted value predicted by the model.

As is clear from Figure \ref{fig:fWvsW} and the discussion above, the high critical density of the HCN J$=$ 1$\rightarrow$ 0 line, 5 $\times 10^5$ cm$^{-3}$ (Shirley 2015), does not necessarily mean that most of the HCN emission from a molecular cloud is emitted by gas at or above that density.  This would certainly be the inference one would draw from our observation that the HCN emission in Perseus covers a significant fraction of the CO emitting area of the cloud, including large portions where A$_K$ $<$ 0.8 mag and where the volume densities are likely below 10$^4$ cm$^{-3}$. As mentioned above, this circumstance has been hinted at in recent wide-field observations of limited portions of a few nearby molecular clouds (e.g., Pety et al. 2017, Shimajiri et al., 2017, Kauffmann et al., 2017) and even some more distant GMCs (Evans et al. 2020, Patra et al. 2022). The mean density of the gas emitting HCN emission in a particular cloud will depend on such factors as the HCN abundance, gas kinetic temperature, mean optical depth and the fraction of cloud mass below the critical density which can still emit HCN sub-thermally.

\begin{figure}
\epsscale{0.8}
\plotone{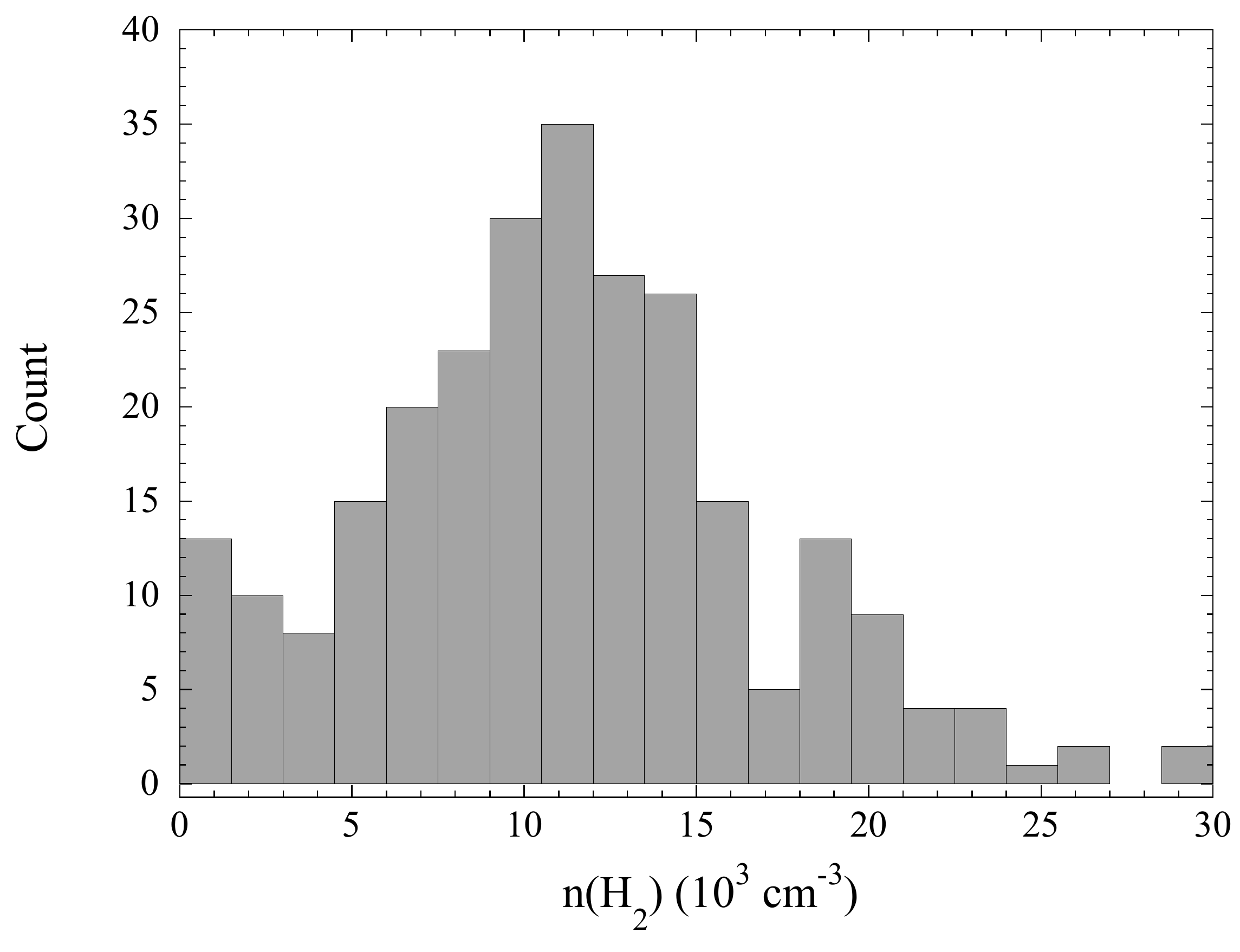}
\caption{The distribution of H$_2$ volume densities at all observed positions, inferred from RADEX modeling of the HCN line intensities.  }
\end{figure}

To make a more quantitative assessment of the mean H$_2$ density exciting the HCN lines, we once more used the radiative transfer program RADEX (van der Tak et al. 2007). However this time we solved for n(H$_2$) using the observed W(HCN) as an input to the calculation. As additional inputs to RADEX we obtained N(HCN) at each position from the dust-derived N(H$_2$), adopting the HCN abundance of 3.1 $\times 10^{-9}$ as above, and again adopted $T_{\rm K}$ $=$ 11 K and  $\Delta$v $=$ 2.3 km s$^{-1}$. The volume density, n(H$_2$), was left as the free parameter, which was varied at each position until W(HCN) matched the observations. The distribution of n(H$_2$) values derived in this way is shown in Figure 7. We find a mean n(H$_2$) of $ 1.1 \pm 0.6 \times 10^4$ cm$^{-3}$, similar to values found by other recent studies cited above. 

\section{Summary}

We have obtained a complete survey of HCN (J=1-0) emission from the nearby Perseus molecular cloud using the CfA 1.2 meter telescope. We have compared our survey with previous complete surveys of CO and dust emission with the following results. 
\\
\\
\noindent
1. The HCN (J=1-0) emission is detected over much of the cloud as defined by complete CO and dust extinction maps. Specifically, HCN is detected at the 3-sigma level over 60\% of the surveyed area. Within this area 71\% of the mass is found to be HCN luminous.  
\\
\\
\noindent
2. Whereas the CO line clearly saturates at N(H$_2$) $\ge$ 4 x 10$^{21}$ cm$^{-2}$, we find that the HCN line remains linear with N(H$_2$) throughout the cloud (up to N(H$_2$) $\sim$ 23 x 10$^{21}$ cm$^{-2}$) and this results in an HCN-to-H$_2$ conversion factor, X$_{HCN}$, of 102 x 10$^{20}$ cm$^{-2}$ K$^{-1}$ km$^{-1}$ s.
\\
\\
\noindent
3. We use a simple RADEX radiative transfer model to demonstrate that, despite its large line optical depths ($\tau$ $=$ 2-25), HCN emission will linearly scale with extinction in filamentary structured molecular clouds where a positive correlation exists between N(H$_2$) and n(H$_2$). To match our HCN observations the model also requires a systematic (exponential) decrease in HCN abundance with increasing extinction, likely due to HCN depletion onto grains. Similar RADEX modeling of the HCN line intensities imply that the mean H$_2$ volume density exciting the HCN lines is $ 1.1 \pm 0.6 \times 10^4$ cm$^{-3}$, a factor of $\sim$50 less than the line's critical density. 
\\
\\
\noindent
4. We obtained the first robust calibration of the dense gas to HCN luminosity ratio in a nearby molecular cloud, $\alpha$(HCN) $=$ 92 M$_\odot$ $(\rm { K\ km\ s^{-1} pc^2})^{-1}$. This is $\sim$9 times higher than the widely-used estimate of Gao \& Solomon. 
\\

\noindent

\bigskip
We gratefully acknowledge financial support from the Scholarly Studies Program of the Smithsonian Institution. We thank the anonymous referee for thoughtful comments that led to improvements in both the clarity and scope of the paper. We also thank Neal Evans for useful suggestions on an early version of this paper.

\end{document}